\documentclass{algoritmy}
\usepackage{float}
\usepackage{booktabs}
\usepackage{multirow}
\usepackage{enumitem}
\usepackage{url}
\usepackage{todonotes}
\usepackage{comment}
\usepackage{threeparttable}
\usepackage{verbatim}

\usepackage{amsmath}
\usepackage{bbding}
\usepackage{multirow} 
\usepackage{graphicx}
\usepackage{amsfonts}
\usepackage{xcolor}
\usepackage{ulem}
\usepackage[misc]{ifsym}
\usepackage[colorlinks,linkcolor=blue,citecolor=blue,urlcolor=blue]{hyperref}

\newcommand{\bR}{\mathbb R}

\newcommand{\bM}{\mathbb M}

\newcommand{\cL}{\mathcal L}

\newcommand{\cA}{\mathcal A}
\newcommand{\cP}{\mathcal P}
\newcommand{\kS}{\mathfrak S}
\usepackage{graphicx}

\title{GAPNet: Granularity Attention Network with Anatomy-Prior-Constraint for Carotid Artery Segmentation}

\author{Lin Zhang\thanks{Shandong Artificial Intelligence Institute, Qilu University of Technology (Shandong Academy of Sciences), Jinan, China.}
\and Chenggang Lu\thanks{School of Artificial Intelligence, Hebei University of Technology, Tianjin, China.}
\and Xin-yang Shi\thanks{School of Computer, Electronics and Information, Guangxi University, Nanning, China.}
\and Caifeng Shan \thanks{School of Intelligence Science and Technology, Nanjing University, Nanjing, China.}
\and Jiong Zhang\thanks{Institute of Biomedical Engineering, Ningbo lnstitute of Materials Technology and Engineering,  Chinese Academy of Sciences} \and Da Chen\thanks{Shandong Artificial Intelligence Institute, Qilu University of Technology (Shandong Academy of Sciences), Jinan, China. ({\tt dachen.cn@hotmail.com})}
\and Laurent D. Cohen\thanks{University Paris Dauphine, PSL Research University, CNRS, UMR 7534, CEREMADE, 75016 Paris, France. ({\tt cohen@ceremade.dauphine.fr})}
 }

\begin{document}


\maketitle

\begin{abstract}
Atherosclerosis is a chronic, progressive disease that primarily affects the arterial walls. It is one of the major causes of cardiovascular disease. Magnetic Resonance (MR) black-blood vessel wall imaging (BB-VWI) offers crucial insights into vascular disease diagnosis by clearly visualizing vascular structures. However, the complex anatomy of the neck poses challenges in distinguishing the carotid artery (CA) from surrounding structures, especially with changes like atherosclerosis. In order to address these issues, we propose GAPNet, which is a consisting of a novel geometric prior deduced from
an anatomical viewpoint. The use of anatomical prior allows the model to avoid segmentation contours whose topology violates the anatomical reality. Specifically, at the first stage, regional features are learned to identify the location of the target CA and to  reduce the influence from the surrounding similar tissues. The second stage aims to improve the feature representation capability, by employing a delicately designed Feature Refinement Attention (FRA) module to capture boundary and detailed information alongside a new Multi-Scale Information Enhancement (MIE) module at the end of the decoder procedure. Experimental results demonstrate the superior performance of our approach on two carotid artery datasets, respectively achieving Dice scores of $0.76$ and $0.83$, proving the effectiveness of GAPNet in improving the accuracy of carotid artery segmentation in MR imaging. 
\end{abstract}


\pagestyle{myheadings}
\thispagestyle{plain}
\markboth{Lin Zhang, Chenggang Lu and Xin-yang Shi}{ALGORITMY 2024 MACRO EXAMPLES}

\section{Introduction}
\label{sec-introduction}
\begin{figure*}[!t]
\centering{\includegraphics[width=0.98\linewidth]{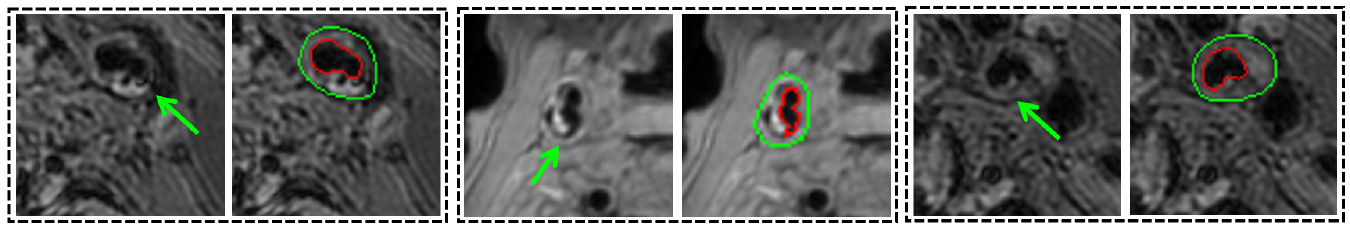}}
\caption{Typical challenges in CA segmentation task from MRI images. The green arrows point to the vessel walls undergoing complex deformations due to lesions.}
\label{fig_VWI}
\end{figure*}

Cardiovascular disease gets to be one of the leading causes of death globally~\cite{tsao2022heart}. Atherosclerosis is a chronic and progressive cardiovascular disease characterized by forming plaques in the arterial intima. These plaques can lead to arterial stenosis, hardening, and plaque rupture, which in turn  cause serious complications such as thrombosis, myocardial infarction, and stroke. Therefore, regular examination of the carotid arteries and early detection of carotid atherosclerosis are essential. Magnetic Resonance (MR) black-blood vessel wall imaging (BB-VWI) can effectively display both normal and pathological arterial vessel walls and characterize atherosclerosis, providing important evidence for clinical diagnosis~\cite{chen2022carotid}. In clinical practice, manual carotid artery segmentation is time-consuming, subjective, and requires specialized training in vessel wall review~\cite{chen2022carotid}. In addition, the complex geometric structures of atherosclerotic lesions and carotid artery  (as shown in Fig.~\ref{fig_VWI}) are  also regarded as a crucial reason that yields difficulties for accurate segmentation. 

In carotid artery segmentation, traditional methods, such as variational models~\cite{seabra20093,yang2012segmentation,ukwatta2011three,hossain2015semiautomatic,chen2019minimal}, usually require specialized domain knowledge, leading to  poor generalization. With the development of fully convolutional networks, many UNet-based methods have emerged. Menchón-Lara et al.~\cite{menchon2016early} employed a perceptron network segment ultrasound CA images. Shin et al.~\cite{shin2016automating} utilized convolutional neural networks (CNNs) to segment ultrasound CA images. Alblas et al.~\cite{alblas2022deep} treated vessel wall segmentation as a multitask regression problem in polar coordinates, encouraging to find  continuous and complete segmented structure of the vessel wall. Despite having improved the accuracy and efficiency of the solutions to the CA segmentation task, most of these methods rely on learning semantic features from images for segmentation. As an important shortcoming, the lack of geometric constraints, suffered by these methods, leads to unacceptable structural errors. Azzopardi et al.~\cite{azzopardi2020bimodal} proposed a geometrically constrained CNN and used amplitude and phase congruency data as input. It imposes shape constraints only considering convex shapes, while the CA may exhibit a certain degree of concavity and convexity simultaneously in reality, especially at bifurcations.

In this paper, we propose a novel granularity attention network and a penalty term based on geometric prior from an anatomical viewpoint, also referred to as anatomical prior, for CA segmentation. A core for the prior lies at the isoperimetric theorem which reveals the essential relation between the perimeter and area of a connected region. More specifically, we design a criterion in terms of the isoperimetric theorem to define the admissibility of a segmented structure. The granularity attention network consists of a two-stage segmentation network combined with the FRA module and the MIE module. This network first performs a coarse segmentation of the region, followed by a refinement process to enhance the network's representational capacity. It is designed to better distinguish the CA from other tissues within the complex anatomy of the neck. The penalty term relying on the introduced  prior of the CA imposes constraint to comply with the anatomical structures.

The main contributions are as follows:
\begin{enumerate}[label=(\alph*)]
    \item We propose a granularity attention network optimized with anatomical prior constraint. The network and constraint ensure the completeness and accuracy of the segmentation by utilizing anatomical prior of the CA and performing feature extraction from coarse to fine granularity.
    \item A novel penalty term is proposed based on the anatomical prior to reduce structurally unacceptable segmentation errors. This prior that is taken as an efficient geometric constraint is able to encourage  to  detect the correct shapes of the CA.
    \item A granularity attention network with the FRA module and the MIE module is designed to enhance the segmentation accuracy. It captures refinement and multi-scale features through a two-stage network for coarse-to-fine-grained segmentation.
\end{enumerate}

\section{Method}
\label{sec-method}
\subsection{Granularity Attention Network}

\begin{figure*}[!t]
    \centering
    \includegraphics[width=1\linewidth]{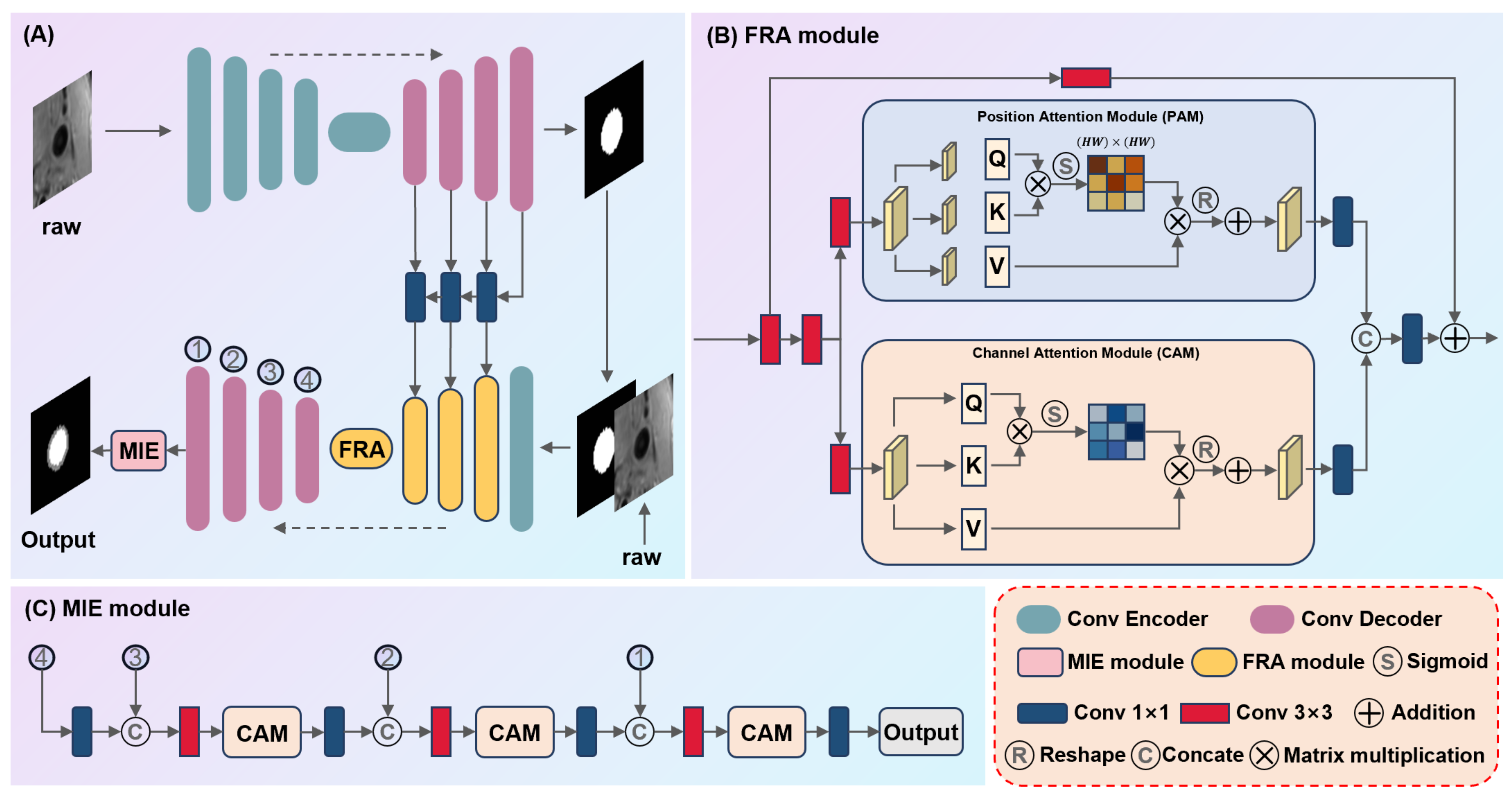}
    \caption{Diagram of the proposed GAPNet.The backbone of the network consists of two U-shaped networks embedded with the FRA module and the MIE module.}
    \label{fig_architecture}
   
\end{figure*}

The complex anatomy of the neck region, where various tissues such as blood vessels, nerves, and soft tissues are densely packed and closely positioned. To precisely segment the CA from complex structures, we propose a granularity attention network based on a two-stage architecture, as shown in Fig.~\ref{fig_architecture}A. The backbone of the network consists of two U-shaped networks. The first stage is employed to extract coarse-grained target regions. It mainly focuses on identifying the approximate contours of the target, performing a rough segmentation of the target area in the image. The decoder output feature maps of the first-stage network are passed through $1\times 1$ convolutions for multi-layer feature fusion before being transmitted to the second-stage encoder. These feature maps contain rich high-level information, and this feature-sharing approach helps the network better understand contextual information, thereby improving segmentation accuracy. The second stage is employed to refine the CA wall and lumen segmentation. The second stage segmentation builds on the first stage segmentation, focusing more on refining the details within the target region. The second stage uses the target area information provided by the first stage to concentrate on more precise segmentation, distinguishing the CA wall and lumen within the target region. In the second stage, we embed the feature refinement attention (FRA) module and the multi-scale information enhancement (MIE) module. The FRA module is employed to refine the higher-level features. The MIE module is used to aggregate information across multiple scales.

The FRA module (as shown in Fig.~\ref{fig_architecture}B) first performs initial feature extraction and aggregation using two $3\times 3$ convolutions on different layers of features from the first stage decoder. Then, we employ position attention and channel attention modules~\cite{fu2019dual} to weight the features with attention, enabling the network to focus on the channels and spatial positions of interest adaptively. Next, we integrate features from channel attention and position attention, leveraging the strengths of both attention mechanisms to enhance feature representation and discrimination capabilities. Finally, we use $1\times 1$ convolutions to adjust the number of parameters, making the model more lightweight and improving computational efficiency and speed. Additionally, through residual connections~\cite{he2016deep}, the original input information is preserved, enhancing gradient flow and improving network training effectiveness. The FRA module can effectively promote cross-layer and cross-stage information interaction. The MIE module (as shown in Fig.~\ref{fig_architecture}C) integrates a channel attention module and aggregates multi-scale features from the decoder. It is employed to enhance the reconstruction of information. Specifically, we use $3\times 3$ convolutions to progressively fuse hierarchical features from different decoder layers, fully utilizing multi-scale information to improve the accuracy of segmentation details. Then, channel attention is applied to enhance important feature information, increasing the model's ability to capture critical information. Additionally, feature dimensions are adjusted through $1\times 1$ convolutions during multi-scale feature aggregation, reducing model complexity and computational cost.

\subsection{Penalty Terms from the Anatomical Prior}

In medical image segmentation, the anatomical prior penalty term refers to a technique that penalizes the segmentation results based on the prior knowledge of known anatomical structures. This penalty term is typically used to guide the segmentation algorithm to follow the known anatomical structures or biological rules when generating segmentation results, thereby improving the accuracy and interpretability of the segmentation. Standard loss functions, such as CrossEntropy loss or Mean Squared Error loss, typically compare the output to the ground truth and quantify their differences. These loss functions usually focus on overall matching between predictions of the model and ground truth. In contrast, anatomical prior constraints provide richer information, aiding the model in better understanding the image content. In this study, we incorporate anatomical prior knowledge of the CA as constraints by adding it as an additional term to the loss function. This additional term penalizes inconsistencies between the output and the anatomical prior, thereby enhancing the robustness and accuracy of the model. The definition of the additional term is as follows.

\begin{figure}[t]
\centering
\includegraphics[width=0.8\textwidth]{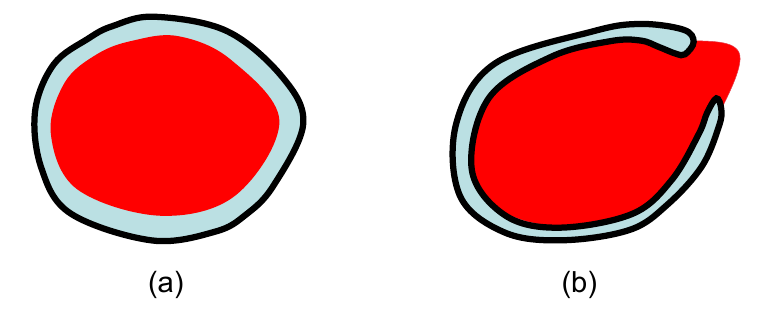}
\caption{\textbf{a} and \textbf{b} respectively illustrate the normal and abnormal segmentation results CA. The regions indicated by cyan color are the segmented CA walls whose external boundary contours are indicated by black solid lines.}
\label{fig_illustrateShapes}
\end{figure}

\noindent \textbf{Topology Prior}:
The CA vessel walls in MRI images usually appear as a \emph{narrow closed band-shape}. However, most of the existing CA segmentation approaches often suffer from the anatomically incorrect leaking problem, where the segmented walls are broken and non-closed.
To address this issue on topology changes, we introduce a novel geometric penalization term based on the \emph{isoperimetric theorem}, encouraging a closed narrow band wall structure. 

We denote by  $\xi:\bM\to[0,1]$  the segmentation prediction of the introduced model, where $\bM\subset\bR^2$ is the open bounded image domain. The segmentation region $\kS\subset\bM$ can be recovered  from the prediction   $\xi$ and a thresholding value $\lambda\in(0,1)$ such that $\xi(x)>\lambda$ means that the point $x$ is inside the segmentation region $\kS$, i.e. $x\in\kS$, and outside $\kS$, otherwise. Let $\chi_\xi$ be a binary function associated with the thresholding value $\lambda$, which reads as
\begin{equation*}
\chi_\xi(x)=
\begin{cases}
1,&\text{if~}\xi(x)>\lambda\\
0,&\text{otherwise}.	
\end{cases}	
\end{equation*}
In this case, the length of the boundary $\partial\kS$ can be denoted by 
\begin{equation*}
\cL(\partial\kS)=\int_\bM \|\nabla\chi_{\xi}(x)\|dx.
\end{equation*}
Moreover, the area of the region $\kS$ reads as
\begin{equation*}
\cA(\kS)=\int_\bM \chi_{\xi}(x)dx.
\end{equation*}
The isoperimetric theorem states that the length $\cL(\partial\kS)$ and the area $\cA(\kS)$ obey $\cL(\partial\kS)^2\geq 4\pi\cA(\kS)$. It measures the relationship between the perimeter of a closed curve and the area it encloses. Therefore,  in terms of the isoperimetric theorem, we define a ratio for measuring the shape of the segmentation region  as follows:
\begin{equation}
\mu(\kS):=\frac{4\pi\cA(\kS)}{\cL(\partial\kS)^2}.
\end{equation}
One can see that the ratio $\mu(\kS)\in\,[0,1]$. In particular, if the region $\kS$ is close to a disk, the ratio  $\mu(\kS)\approx 1$, while when the segmentation region $\kS$ appears as a \emph{non-circular narrow closed band} shape, $\mu(\kS)$  gets to be small. Specific to the CA vessel wall segmentation task, the leaking problem usually yields a narrow band shape of \emph{strong concavity}, as shown in Fig.~\ref{fig_illustrateShapes}(b), which leads to a very low value of $\mu(\kS)$. For the normal case, the segmented CA vessel walls can be approximately delineated via either a disk-like shape, or an elliptical-like shape, or a union of multiple disk-like shapes, leading to $\mu(\kS)\approx1$. 

Providing that the boundary $\partial\kS$  represents the \emph{external} contour of the segmented CA vessel wall structure. We define that a segmented CA vessel wall is admissible if the ratio $\mu(\kS)>\tau$.  
Specifically, we consider the following penalization term 
\begin{equation}
\label{eq_penaltyMorphological}
\cP_{\rm topology}:= \max \left\{0,-(\mu(\kS) - \tau) \right\},
\end{equation}
where $\tau$ is a thresholding value and is set as $\tau=0.6$ through this paper. When $\mu(\kS) > \tau$, the value of $\mu(\kS) - \tau$ is positive, indicating that the topological structure of the segmentation region is acceptable. In this case, taking the negative of  $\mu(\kS) - \tau$ results in a value less than $0$, ensuring that no penalty is applied in this case. For segmentation results that are below the thresholding value $\tau$, the value of $\mu(\kS) - \tau$ is less than 0. After taking the negative of  $\mu(\kS) - \tau$, the result is greater than zero, applying a penalty to these abnormal segmentation results, thus encouraging the model to generate satisfactory segmentation predication.

\section{Experimental Results}
\subsection{Dataset}
We used data from the Carotid Vessel Wall Segmentation and Atherosclerosis Diagnosis Challenge (COSMOS 2022)~\cite{chen2022carotid} and the Carotid Artery Vessel Wall Segmentation Challenge 2021(CAVWSC 2021)~\cite{zhao2017chinese} to validate our method. The COSMOS 2022 dataset consists of 50 3D MR scans. We randomly split 40 cases ($80\%$) for training and validation, reserving 10 cases ($20\%$) for testing. Among the 40 cases used for training and validation, we obtained 934 annotated slices, with $80\%$ randomly assigned for training and $20\%$ for validation. The testing data comprised 268 annotated slices. For the CAVWSC dataset, we obtained 1737 annotated slices for training and validation, with $80\%$ randomly assigned for training and $20\%$ for validation. The testing set contained 1754 annotated slices. Each image was resized to a resolution of $224 \times 224$ for experimentation.

\subsection{Implementation Details}
Our method uses the PyTorch framework, and experiments are conducted on NVIDIA Tesla V100. We employ the AdamW optimizer with a base learning rate of $0.0001$. In the first stage, the network uses Dice loss and cross-entropy loss to supervise the region segmentation; in the second stage, the network uses a combination of Dice loss, cross-entropy loss, and penalty term to supervise the target segmentation. For the first $5$ epochs, we utilize the Warm-up learning rate strategy, followed by the Polynomial Decay strategy for learning rate decay after $5$ epochs. The batch size is set to $8$, and we train for $300$ epochs. Data augmentation is performed by adding brightness and Gaussian noise and applying random scaling, rotation, shifting, and cropping. We adopt the Dice coefficient (Dice) and Hausdorff distance (HD) as evaluation metrics. A 5-fold cross-validation is used to evaluate the model.

\begin{figure*}[!t]
\centering
\includegraphics[ width=1\linewidth]{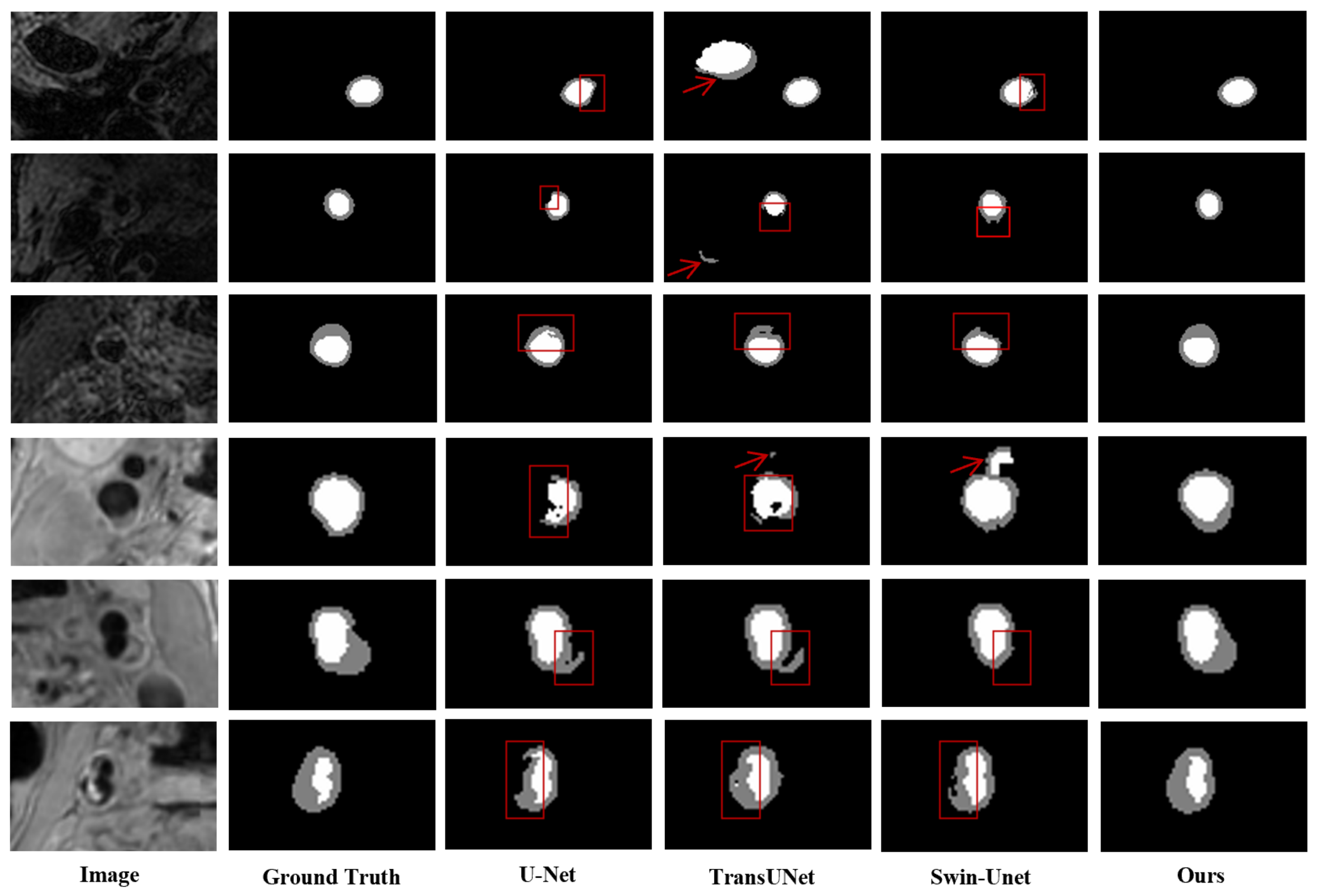}
\vskip -5pt
\caption{Comparison of different methods. Arrows indicate excessive segmentation and boxes denote incomplete segmentation structures or insufficient details.}
\label{fig_compare}
\end{figure*}

\begin{table*}[htbp]
\setlength{\tabcolsep}{5pt}
\renewcommand{\arraystretch}{1.5}
\centering
\caption{Performance comparisons for vessel wall and lumen segmentation.}
\resizebox{\textwidth}{!}{
    \begin{threeparttable}
        \begin{tabular}{l|cccc|cccc}
        \toprule
        \multicolumn{1}{c}{\multirow{3}{*}{Methods}} & \multicolumn{4}{c}{CAVWSC 2021} & \multicolumn{4}{c}{COSMOS 2022} \\ \cmidrule{2-9}
        \multicolumn{1}{c}{~} & \multicolumn{2}{c}{Vessel Wall} & \multicolumn{2}{c}{Lumen} & \multicolumn{2}{c}{Vessel Wall} & \multicolumn{2}{c}{Lumen} \\ 
        \multicolumn{1}{c}{~} & Dice $\uparrow$ & HD $\downarrow$ & Dice $\uparrow$ & HD $\downarrow$ & Dice $\uparrow$ & HD $\downarrow$ & Dice $\uparrow$ & HD $\downarrow$ \\ \midrule
        U-Net\,\cite{ronneberger2015u} &0.7295 &10.5739 &0.9154 &9.7920 &0.8054 &5.6000 &0.9152 &4.8744 \\
        UNet++\,\cite{zhou2018unet++} &0.7432 &10.3248 &0.9130 &9.3279 &0.8226 &4.7868 &0.9213 &4.3572 \\
        Attention U-Net\,\cite{oktay2018attention} &0.7415 &9.5014 &0.9184 &8.7724 &0.8214 &4.8154 &0.9246 &5.8245 \\
        DualAttentionU-Net\,\cite{yu2020dual} &0.7356 &7.5142 &0.9203 &7.9179 &0.8232 &4.0071 &0.9297 &3.5967 \\
        Res-UNet\,\cite{xiao2018weighted} &0.7473 &8.3080 &0.9190 &7.9907 &0.8212 &3.8654 &0.9266 &3.6677 \\ 
        TransUNet\,\cite{chen2021transunet} &0.7528 &7.4204 &0.9223 &7.1371 &0.8209 &4.4920 &0.9295 &5.5275 \\ 
        Swin-Unet\,\cite{cao2022swin} &0.7485 &6.3345 &0.9160 &4.3895 &0.8050 &3.6810 &0.9200 &3.1113 \\ 
        \textbf{Proposed} & \textbf{0.7666} & \textbf{4.6540} & \textbf{0.9328} & \textbf{3.5642} & \textbf{0.8397} & \textbf{2.7375} & \textbf{0.9345} & \textbf{2.6059} \\ 
        \bottomrule
        \end{tabular}
        \label{tab_comparison}
    \end{threeparttable}
}
\end{table*}

\subsection{Comparison with State-Of-The-Art Models}
We compared the proposed method with seven advanced medical image segmentation methods, namely U-Net~\cite{ronneberger2015u}, UNet++~\cite{zhou2018unet++}, Attention U-Net~\cite{oktay2018attention}, Dual Attention U-Net~\cite{yu2020dual}, Res-UNet~\cite{xiao2018weighted}, TransUNet~\cite{chen2021transunet}, and Swin-Unet~\cite{cao2022swin}. We conducted the same training and testing on both the COSMOS and CAVWSC datasets using these methods, and the results are shown in Table~\ref{tab_comparison}. Our method performs optimally in Dice and HD evaluation metrics for segmenting the vessel wall and lumen. Specifically, for the challenging task of vessel wall segmentation, our method achieves a Dice of $0.7666$ and an HD of $4.6540$ on the CAVWSC  dataset and a Dice of $0.8397$ and an HD of $2.7375$ on the COSMOS dataset.

Fig.~\ref{fig_compare} exhibits the qualitative comparison results that depict the CA in MRI images. The boxes highlight erroneous segmentations caused by complex diseased vessel wall structures or noise interference in the images, producing discontinuities in the results that do not align with the real anatomical structure. On the other hand, the red arrows indicate that TransUNet and Swin-Unet mistakenly identify tissues with similar features as the CA. In contrast, our method excels in finely and accurately segmenting vessel wall structures, yielding smoother segmentation contours.

\begin{figure*}[!t]
\centering
\includegraphics[width=0.95\linewidth]{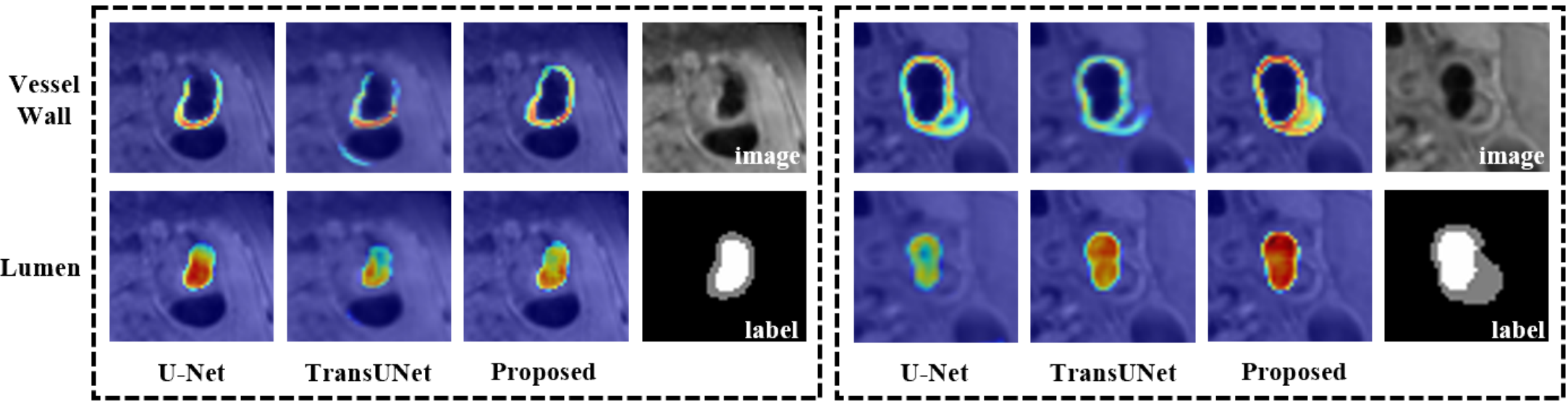}
\vskip -5pt
\caption{Visualization of the heat map from the final layer of the decoder.}
\label{fig_heatmaps}
\end{figure*}

The heatmap in Fig.~\ref{fig_heatmaps} further confirms that our network can focus on more complete vessel wall areas, particularly in complex diseased vessel wall structures. In summary, our method excels in accurately locating targets in complex images, overcoming interference from nearby similar tissues, and producing more detailed and precise segmentation results, thereby  significantly improving result accuracy.


\begin{table*}[htbp]
\setlength{\tabcolsep}{6pt}
\renewcommand{\arraystretch}{1.5}
\centering
\caption{Ablation  results for vessel wall and lumen segmentation.}
\resizebox{\textwidth}{!}{
    \begin{threeparttable}
        \begin{tabular}{l|cccc|cccc}
        \toprule
        \multicolumn{1}{c}{\multirow{3}{*}{Methods}} & \multicolumn{4}{c}{CAVWSC 2021} & \multicolumn{4}{c}{COSMOS 2022} \\ \cmidrule{2-9}
        \multicolumn{1}{c}{~} & \multicolumn{2}{c}{Vessel Wall} & \multicolumn{2}{c}{Lumen} & \multicolumn{2}{c}{Vessel Wall} & \multicolumn{2}{c}{Lumen} \\ 
        \multicolumn{1}{c}{~} & Dice $\uparrow$ & HD $\downarrow$ & Dice $\uparrow$ & HD $\downarrow$ & Dice $\uparrow$ & HD $\downarrow$ & Dice $\uparrow$ & HD $\downarrow$ \\ \midrule
        w/o Stage1+M1 &0.7515 &7.0755 &0.9222 &6.5112 &0.8168 &4.0999 &0.9258 &3.6526 \\
        w/o Stage1 &0.7554 &6.3529 &0.9202 &5.7233 &0.8194 &3.9221 &0.9273 &3.7459 \\
        Backbone &0.7569 &5.8109 &0.9154 &5.4130 &0.8243 &3.7496 &0.9253 &3.1835 \\
        Backbone+M1 &0.7622 &5.9202 &0.9269 &5.3959 &0.8276 &3.2927 &0.9259 &2.8477 \\
        Backbone+M2 &0.7636 &5.2730 &0.9290 &4.7746 &0.8351 &3.2658 &0.9321 &2.8156 \\
        Backbone+M3 &0.7605 &5.6511 &0.9268 &4.7736 &0.8257 &3.2222 &0.9232 &2.7375 \\ 
        Backbone+M1+M2 &0.7628 &5.3616 &0.9288 &4.6485 &0.8328 &3.0883 &0.9300 &2.5990 \\ 
        Backbone+M1+M3 &0.7633 &5.3028 &0.9273 &4.5919 &0.8287 &3.4332 &0.9223 &2.9572 \\ 
        Backbone+M2+M3 &0.7632 &5.1912 &0.9258 &4.4615 &0.8358 &3.0950 &0.9343 &\textbf{2.4486} \\
        Backbone+M1+M2+M3 & \textbf{0.7666} & \textbf{4.6540} & \textbf{0.9328} & \textbf{3.5642} & \textbf{0.8397} & \textbf{2.7375} & \textbf{0.9345} & 2.6059 \\ 
        \bottomrule
        \end{tabular}
        \label{tab_ablation}
    \end{threeparttable}
}
\end{table*}

\subsection{Ablation Study}
To validate the effectiveness of our proposed methods M1: anatomical prior constraint, M2: FRA module, and M3: MIE module in the CA segmentation task, we utilized a two-stage granularity network as the backbone and gradually incorporated our methods for ablation experiments. As shown in Table~\ref{tab_ablation}, the experimental results indicate that adding our proposed methods individually to the backbone, or combining them in pairs in different ways, all had a positive impact on the experimental results. Finally, the comprehensive GAPNet integrating M1, M2, and M3 achieved the best overall performance on both datasets. On the other hand, to verify the effectiveness of the first stage in extracting coarse-grained target regions for this task, we conducted experiments by removing the first stage network separately. The experimental results showed that, compared to GAPNet, removing the first stage coarse-grained extraction network had a negative impact on the experimental results. On the CAVWSC dataset, the Dice coefficients for segmenting the vessel wall and lumen decreased by $1.12\%$ and $1.26\%$, respectively; on the COSMOS dataset, they decreased by $2.03\%$ and $0.72\%$, respectively. All metrics on both datasets showed a significant decline, indicating that the first stage coarse-grained extraction network helps the network accurately locate target regions, achieving more precise segmentation. In experiments removing the first stage coarse-grained extraction network, we further validated the effectiveness of M1 through ablation. As shown in Table~\ref{tab_ablation}, the results indicate that introducing M1 positively impacts the model's performance, especially in the vessel wall region. This demonstrates the effectiveness of anatomical priors for carotid vessel segmentation.

\section{Conclusion and Future Work}
In this work, we introduce an effective method for fully automated and precise segmentation of CA in MRI. Our approach introduces a novel granularity attention network, enabling segmentation from coarse to fine-grained levels and enhancing the ability to capture boundary and detail information through the FRA module and the MIE module. Additionally, anatomical prior constraints are introduced to adjust the segmentation results, thereby improving segmentation completeness and accuracy. Comprehensive experimental results demonstrate that our method achieves excellent performance on two publicly available datasets, further demonstrating the accuracy and robustness of the model segmentation.

We note that the proposed segmentation model indeed does not take into account geometric regularization such as curvature-based length, and more types of  shape priors, for instance the star convexity shape prior which is  an important cue for defining the expected segmentation contours in the CA segmentation task. Future work will be devoting to solving these limitations.

\section*{Acknowledgments}
This work is in part supported by the National Natural Science Foundation of China (62371442, 62103398), the Shandong Provincial Natural Science Foundation (NO.~ZR2022YQ64), the Natural Science Foundation of Zhejiang Province (LZ23F010002, LR24F010002) and the French government under management of Agence Nationale de la Recherche as part of the ``Investissements d'avenir'' program, reference ANR-19-P3IA-0001 (PRAIRIE 3IA Institute).

\hspace*{\fill} \\
\noindent

\bibliographystyle{splncs}
\bibliography{refs}

\end{document}